\documentclass[prl,twocolumn,showpacs,aps,floatfix,amsmath,amssymb,amsfonts]{revtex4-1}

\usepackage{amsmath}
\usepackage{graphicx}
\usepackage{color}

\usepackage{mathrsfs}
\usepackage{amssymb,amsfonts,mathrsfs}
\usepackage{bm}

\usepackage{bbold}

\newcommand{\beq}{\begin{equation}}
\newcommand{\eeq}{\end{equation}}

\definecolor{ao}{rgb}{0.0, 0.5, 0.0}

\begin{document}

\title{Single-atom transistor as a precise magnetic field sensor}
\author{Krzysztof Jachymski$^{1}$}
\author{Tomasz Wasak$^2$}
\author{Zbigniew Idziaszek$^2$}
\author{Paul S. Julienne$^3$}
\author{Antonio Negretti$^4$}
\author{Tommaso Calarco$^5$}
\affiliation{
$^1$ Institute for Theoretical Physics III \& Center for Integrated Quantum Science and Technologies (IQST), University of Stuttgart, Pfaffenwaldring 57, 70550 Stuttgart, Germany\\
$^2$ Faculty of Physics, University of Warsaw, Pasteura 5, 02-093 Warsaw, Poland\\
$^3$ Joint Quantum Institute, University of Maryland and National Institute of Standards and Technolog, College Park, Maryland 20742, USA\\
$^4$ Zentrum f\"ur Optische Quantentechnologien and The Hamburg Centre for Ultrafast Imaging, Universit\"at Hamburg, Luruper Chaussee 149, 22761 Hamburg, Germany\\
$^5$ Institute for Complex Quantum Systems \& Center for Integrated Quantum Science and Technologies (IQST), Universit\"at Ulm, 89069 Ulm, Germany}

\date{\today}

\begin{abstract}
Feshbach resonances, which allow for tuning the interactions of ultracold atoms with an external magnetic field, have been widely used to control the properties of quantum gases. We propose a~scheme for using scattering resonances as a probe for external fields, showing that by carefully tuning the parameters it is possible to reach a $10^{-5}$G (or nT) level of precision with a single pair of atoms. We show that for our collisional setup it is possible to saturate the quantum precision bound with a simple measurement protocol.
\end{abstract}

\maketitle

{\it Introduction}.
Quantum technologies hold the promise for significant advancement in various fields such as communication and sensing due to the potential to utilize quantum coherence or entanglement to improve the performance of devices. 
In recent years, great progress has been made in bringing quantum-enhanced sensing towards practical and industrial applications~\cite{Giovannetti2006,Degen2016}. Here the goal is to construct specific quantum systems for the precise measurement of external parameters such as electromagnetic fields. This is crucial in a range of domains from fundamental~\cite{Webb1999,Chin2006,Zelevinsky2008,Blatt2008,Schnabel2010,Hudson2011,Baron2014} to technological applications e.g. in medicine or materials science, where detection of fields produced by single spins is often desired~\cite{Balasubramanian2008}.

The state-of-the-art magnetic field sensing techniques exploit field-dependent
effects in a number of different systems. Outstanding sensitivity to ac signals is obtained by the superconducting quantum interference devices (SQUID)
\cite{Jaklevic1964quantum,Vasyukov2013}. Other systems that reach high performance are based on nitrogen-vacancy centres in diamonds~\cite{Maze2008,Wolf2015, Zaiser2016}, thermal atomic
vapors~\cite{Lucivero2014,wasilewski2010quantum}, internal states of trapped ions~\cite{Baumgart2016,kotler2011single}, and the cold or ultracold atomic
samples~\cite{kominis2003subfemtotesla,Wildermuth2005,Vengalattore2007,Koschorreck2011,Behbood2013,Yang2017,Ciurana2017,Peters2017}. Ultracold atoms are a natural candidate for implementing quantum sensing protocols, since they offer the possibility of working with large ensembles of particles prepared in a very well defined initial quantum state. Due to low (sub-$\mu$K) temperatures, cold atoms can be trapped using external electromagnetic fields such as optical lattices~\cite{Bloch2008}. Interestingly, interatomic interactions can also be tuned in an experiment if a Feshbach resonance is available~\cite{Weiner1999,Chin2010}. The resonance mechanism originates from the coupling of the free atomic pair to a bound state. Typically the different scattering channels are associated with the hyperfine structure of the atoms, which allows to tune the position of the bound state via magnetic field by means of the differential Zeeman shift.

\begin{figure}
\centering
\includegraphics[width=0.45\textwidth]{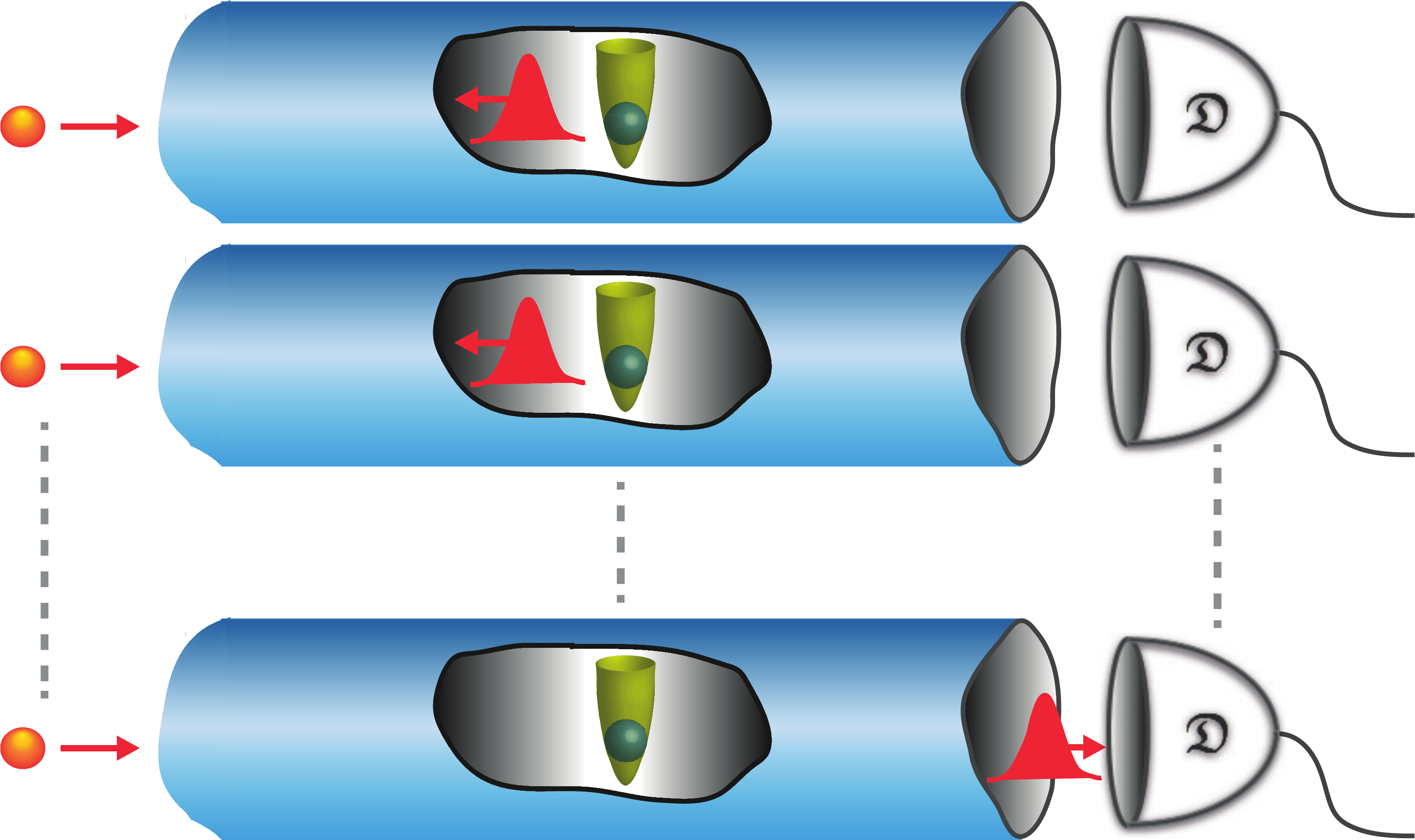}
\caption{\label{fig:setup} (Color online). Sketch of the magnetic field sensor, where $N$ atoms (red spheres on the left) are sent through $N$ quasi-one dimensional waveguides (in the picture  represented by the tubes).
In each waveguide a tightly confined impurity atom (green sphere) is placed. The colliding atoms (red wave-packets) 
can be either transmitted or reflected with probability depending on the total external magnetic field strength.}
\end{figure}

The interplay of controlled interactions and external confinement has been the subject of intense studies, both experimental~\cite{Moritz2005,Haller2010,Sala2013} and theoretical~\cite{Olshanii1998,Tiesinga2000,Petrov2001,Bolda2003,Bergeman2003,Granger2004,Idziaszek2006,Naidon2007,Yurovsky2007,Sala2012,Giannakeas2013,Melezhik2016,Jachymski2017,Greene2017}.
Confinement-induced resonances, which result from the modification of scattering properties by the external trap, allowed for controlling atomic interactions of ultracold bosons in low dimensions, leading to the experimental realization of the long-sought Tonks-Girardeau gas~\cite{Paredes2004,Kinoshita2004}. 

In this Letter, we propose to look at the resonances from a different angle. Instead of tuning the collisional properties of atoms with external fields, we treat the collisions as a probe of the field itself. We consider a simple scheme in which the atoms are colliding in quasi-one-dimensional (1D) waveguides created e.g. by an optical lattice. In the vicinity of a Feshbach resonance, the collisional phase shift strongly depends on the external magnetic field. Information about the field strength can be extracted e.g. by measuring the transmission of atoms through the waveguides. 
The performance of this measurement can be assessed by means of the Quantum Cram\'er-Rao
lower bound (QCRLB), which provides the ultimate limit on the uncertainty of the inferred magnetic field \cite{braunstein1994statistical}.

{\it Sensor construction}.
The sensor we have in mind is schematically illustrated in Fig.~\ref{fig:setup}. An ensemble of $N$ noninteracting atoms is injected into a set of quasi-one-dimensional waveguides realized by using a deep 3D optical lattice relaxed in the longitudinal direction. In the center of every waveguide there is a tightly confined impurity atom, either from a different hyperfine state or different species, tightly trapped by an optical potential using a magic wavelength transparent for the incoming atoms~\cite{Clark2015}. The sensitivity to magnetic fields is provided by the Feshbach resonance, which controls the interaction between the atom and the impurity characterized by the 3D scattering length $a(B)$.
The transverse width of the waveguides $d$ is chosen in such a way that the probability of reflecting the colliding atom back from the impurity strongly depends on the value of the magnetic field, which is explained in detail further. The efficient detection of the transmitted single atoms at the end of the waveguides can be accomplished in several ways, for example by ionization or absorption imaging. From the counting statistics it is then possible to infer the strength of the magnetic field with the precision attaining the classical Cram\'er-Rao bound. As shown in Fig.~\ref{fig:setup}, the spatial separation of the tubes typically of the order of 500~nm naturally provides high spatial resolution. Systems with desired properties can be realized with state-of-the-art  techniques used in ultracold atomic quantum gas experiments (see e.g.~\cite{Meinert2016,Alberti2017}).

{\it Atomic scattering in quasi-one-dimensional geometry}.
Let us now briefly review the relevant two-body physics taking place in a single tube. We assume the impurity atoms are pinned by the trap~\cite{SupMat}. 
The stationary Schr\"{o}dinger equation for the incoming atom then reads $\left[{\bf r}\equiv (x,y,z)\right]$
\beq
\label{eq1}
\left[-\frac{\hbar^2}{2m}\nabla^2+U(\mathbf{r})+V_{\rm tr}(\mathbf{r})\right]\Psi(\mathbf{r})=E\Psi(\mathbf{r}).
\eeq
Here $V_{\rm tr}$ is the transverse trap which we assume to be harmonic $V_{\rm{tr}}=\frac{1}{2}m \omega^2 \rho^2$ ($\rho^2 = x^2 + y^2$), and $\omega$ is the trap frequency. The parameters can be combined into a characteristic lengthscale $d=\sqrt{\hbar/m\omega}$. Finally, $U(\mathbf{r})$ is the interparticle interaction with characteristic range much smaller than $d$ and it can be described by the pseudopotential~\cite{Bloch2008,SupMat}
\beq
U_s(\mathbf{r}) = \frac{2\pi \hbar^2 \tilde{a}(k)}{\mu}\delta(\mathbf{r})\frac{\partial}{\partial r}(r\cdot). \label{s-pseudo}\\
\eeq
Here $\mu$ is the reduced mass of the pair of atoms, and the energy-dependent scattering length is defined as $\tilde{a}(k)=-\tan \delta_{\ell=0}(k)/k$, where $\delta_\ell$ denotes the phase shift in the partial wave $\ell$. The scattering length depends on the magnetic field due to a Feshbach resonance and in the zero energy limit can be described by the simple relation~\cite{Chin2010}
\beq
\tilde{a}(k=0,B)=a_{\rm{bg}}\left(1-\frac{\Delta}{B-B_{\rm res}}\right),
\label{a3d}
\eeq
where $a_{\rm{bg}}$ is the background scattering length away from the resonance, $\Delta$ denotes the resonance width, and $B_{\rm res}$ is the resonance position. In order to work in the incoming atoms reference frame we rewrite eq.~\eqref{s-pseudo} as $U_s=\frac{2\pi \hbar^2 a(k)}{m}\delta(\mathbf{r})\frac{\partial}{\partial r}(r\cdot)$ with $a(k)=\frac{m}{\mu}\tilde{a}(k)$. Here $k$ is the total energy of the relative motion $E=\frac{\hbar^2 k^2}{2\mu}=\hbar\omega+\frac{\hbar^2 p^2}{2m}$, where 
$p$ is the one-dimensional wavenumber.

In the presence of a strong transverse confinement one can assume that the asymptotic wave function is well described by the lowest mode of the transverse harmonic oscillator. The transmission coefficient, which describes the part of the flux that goes through the tube, can then be defined as
$T(p)=\cos^2\delta_{1D}(p)$~\cite{Olshanii1998},
with $\delta_{1D}$ being the one-dimensional phase shift given by~\cite{Bergeman2003}
\beq
\label{delta1D}
p\tan\delta_{1D}(p)=-\frac{2}{d}\left(\frac{d}{a(k)}-C(k)\right)^{-1}.
\eeq
Here $C=-\zeta_H\left(\frac{1}{2},\frac{3}{2}-\frac{E}{2\hbar\omega}\right)$ and $\zeta_H$ is the Hurwitz zeta function.

\begin{figure}
\centering
\includegraphics[width=0.23\textwidth]{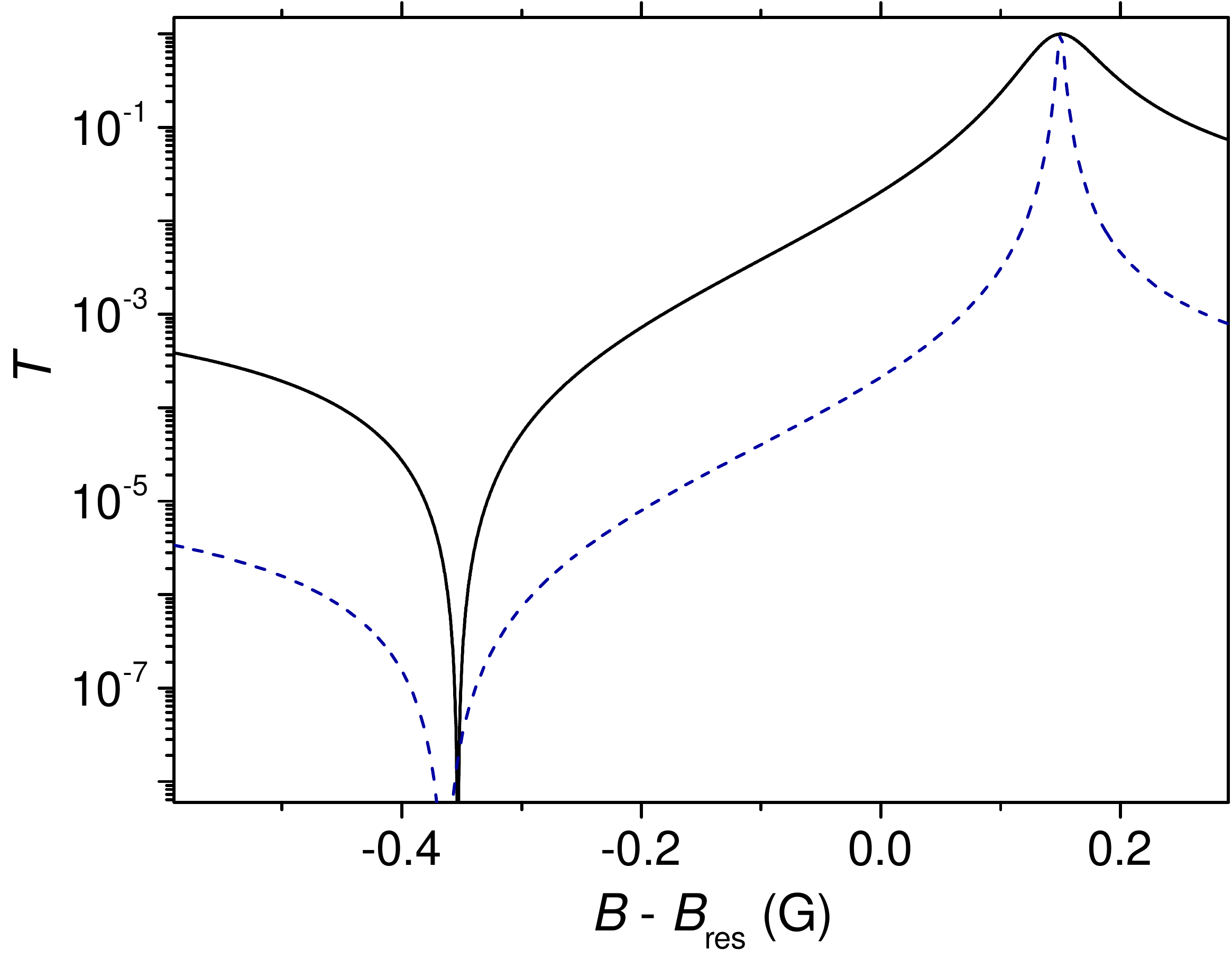}
\includegraphics[width=0.23\textwidth]{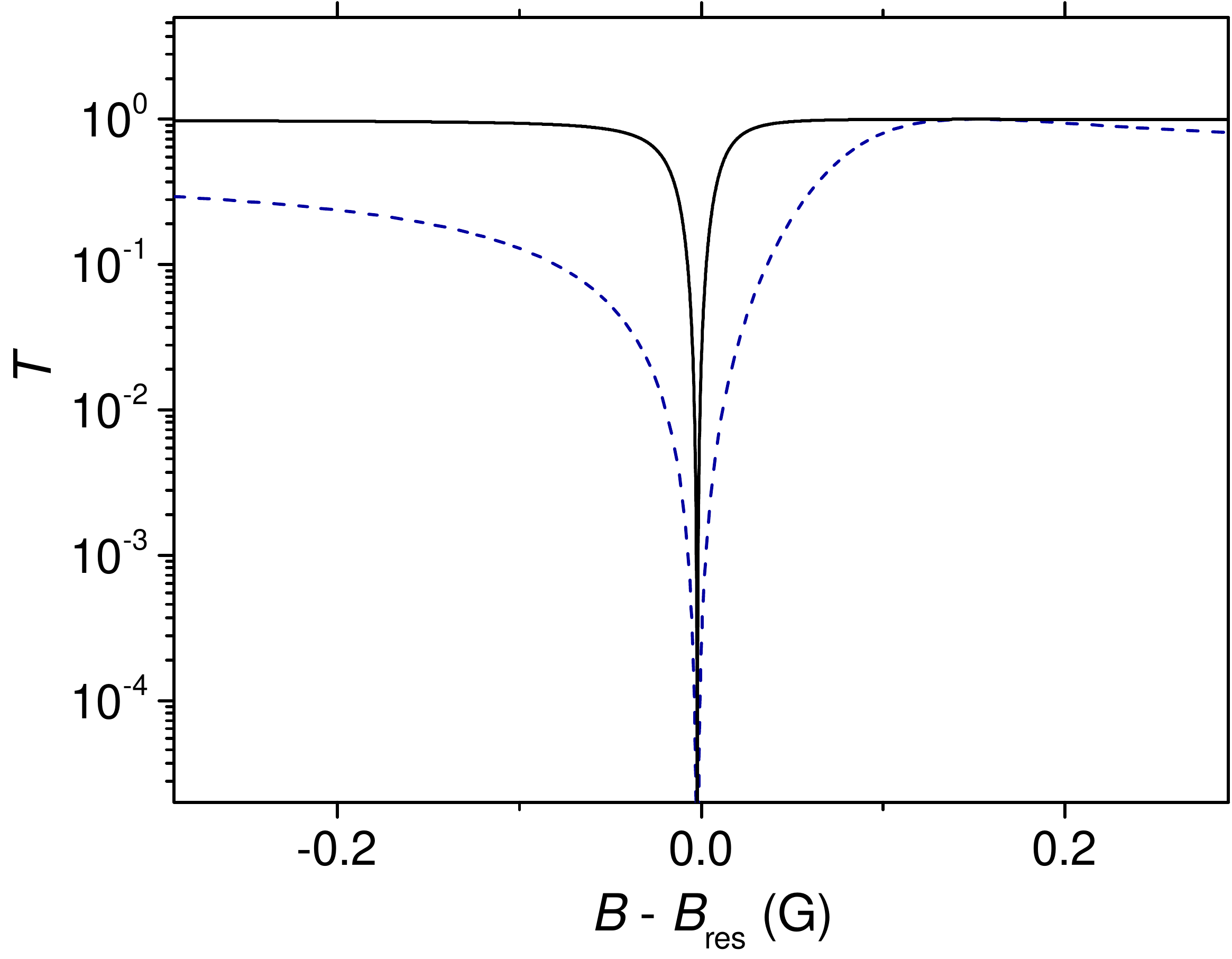}
\caption{\label{fig:transm} Transmission coefficient as a function of the magnetic field for two exemplary resonances with the same width $\Delta=0.15$G, but different
  background scattering lengths: $a_{\rm bg}=9.7\bar{a}$ (left), similar e.g. to Cs atoms, and $a_{\rm bg}=0.2\bar{a}$ (right).  Dashed blue lines show the results for
  very low longitudinal momentum $p=0.001\bar{a}^{-1}$, while for solid black lines $p=0.01\bar{a}^{-1}$, corresponding typically to about 0.1 nK and 15 nK.}
\end{figure}

In Fig.~\ref{fig:transm} we show the dependence of the transmission coefficient on the magnetic field computed for two exemplary Feshbach resonances generated numerically using a two-channel model with van der Waals interactions in a quasi-1D harmonic trap. For the length unit we use $\bar{a}$ defined as the mean characteristic length of the van der Waals potential~\cite{Gribakin}, which is typically of the order of $\sim 100a_0$. Optical lattice confinement leads to values of $d$ of around $15-25\bar{a}$~\cite{Haller2010} while we choose $d=20\bar{a}$. At the position of the so-called confinement-induced resonance (CIR) given by $d/a(B)=C(k) \approx 1.4603$, $\tan\delta_{1D}$ diverges and the transmission reaches zero. One can also observe the opposite case of unit transmission, where the atoms are effectively noninteracting near the zero crossing of the 3D scattering length. Both features can in principle be used for magnetic field measurement. We note that while the CIR feature becomes sharper as the collision energy increases, the unit transmission peak becomes less pronounced. This is because the background transmission grows with energy.

{\it Sensor performance}.
We proceed with the analysis of the achievable sensor performance.
The classical Cram\'er-Rao lower bound (CRLB) is a general theorem from estimation theory that provides a lower bound on the uncertainty of the inferred value of an unknown 
parameter~\cite{braunstein1994statistical, refregier2012noise, cramer2016mathematical}. In our case, this bound can be
expressed in the form of the inequality for the estimation uncertainty $\Delta B_\mathrm{est}$: $(\Delta B_\mathrm{est})^2 \geqslant 1 / (N F)$, where $F$ is the classical Fisher information~\cite{refregier2012noise} which quantifies the usefulness of the metrological protocol and $N$ is the number of atoms. 
The Fisher information is expressed in terms of the probability distribution of different outcomes
\begin{equation}
\label{cfi}
F\!= \! \sum_{s=\pm1} \frac{1}{P(s|B)}\!\! \left(\!\frac{\partial P(s|B)}{\partial B}\!\right)^2,
\end{equation}
where the transmission probability is $P(+1|B)\equiv T(B)$, and the probability of reflecting the atom is $P(-1|B)\equiv 1-T(B)$. 
The CRLB is saturated asymptotically by the maximum likelihood estimator in the limit of a large number of atoms used in the estimation procedure. 

Expressing the probability distributions in terms of the transmission coefficient $T(B)$, the Fisher information takes the  form
\begin{equation}
  F = \frac{1}{T(B)\left(1-T(B)\right)} \left(\frac{d T(B)}{dB}\right)^2.
\label{fishers}
\end{equation}
The structure of this formula is intuitively clear, as the most favourable conditions are attained when the transmission strongly depends on the magnetic field. 
The uncertainty is further reduced by the statistical enhancement factor $\sqrt{NM}$, where 
$M$ denotes the number of repetitions (or atoms per tube). Hence, with a limited experimental effort, one can easily improve the sensor sensitivity by several orders of magnitude.

\begin{figure}
\centering
\includegraphics[width=0.23\textwidth]{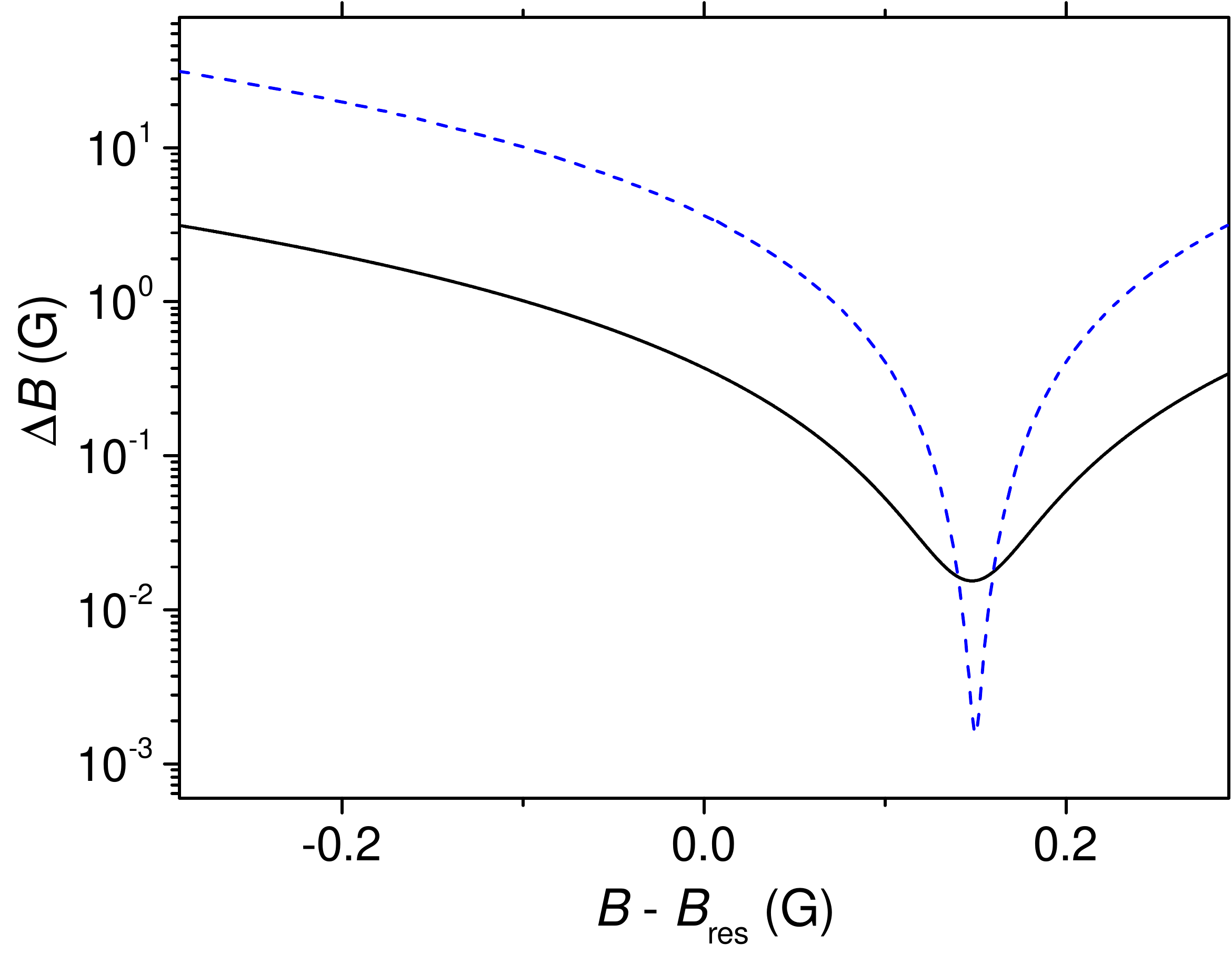}
\includegraphics[width=0.23\textwidth]{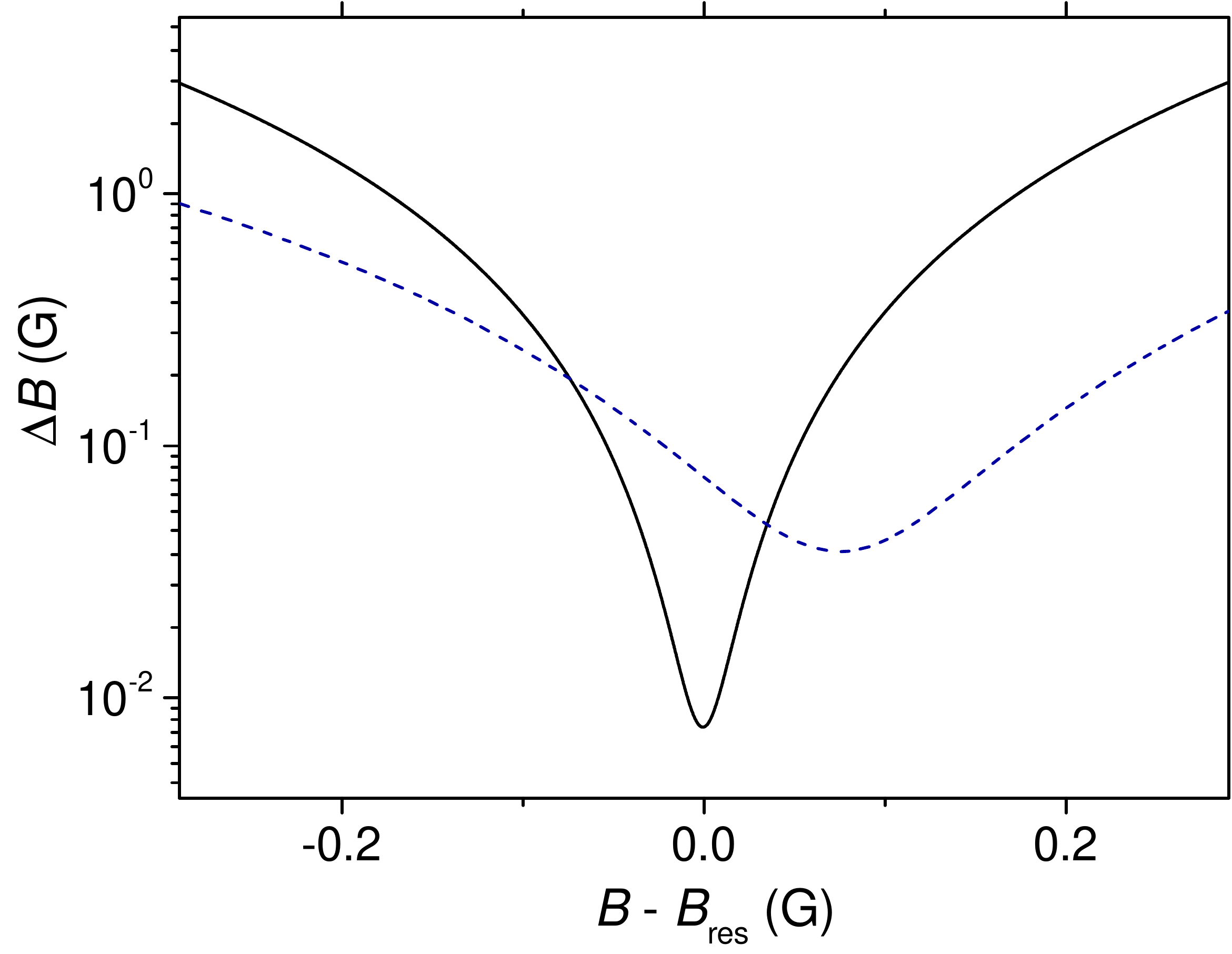}
\caption{\label{fig:prec}Maximum achievable precision in a single shot ($N=M=1$) as calculated from Eq.~\eqref{fishers} for the same parameters as in Fig.~\ref{fig:transm}.}
\end{figure}

\begin{figure*}[t]
\centering
\includegraphics[width=0.28\textwidth]{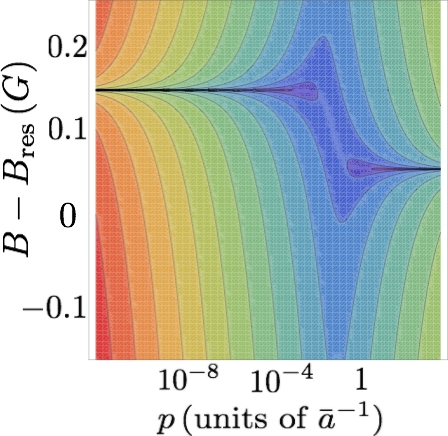}
\includegraphics[width=0.28\textwidth]{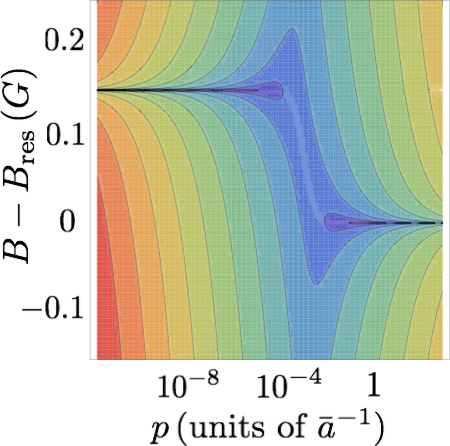}
\includegraphics[width=0.08\textwidth]{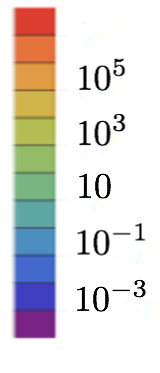}
\includegraphics[width=0.33\textwidth]{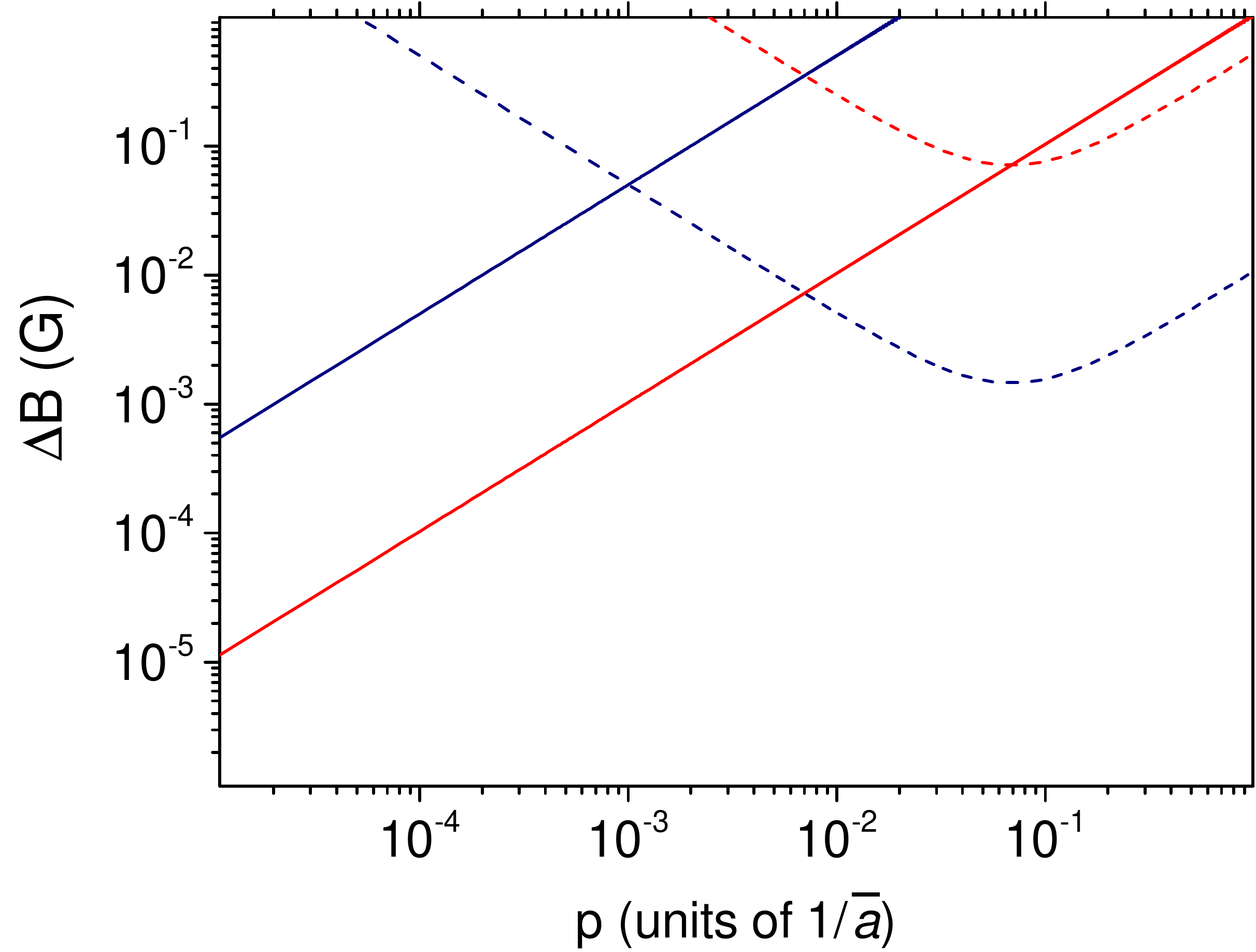}
\caption{\label{fig:precdelta}The best achievable precision $\Delta B$ in Gauss units calculated from Eq.~\eqref{fishers} (neglecting the finite energy corrections to the scattering phase shift) as a function of the field value and longitudinal momentum $p$ for $\Delta=0.15$G. Left: high background scattering length $a_{\rm bg}=9.7\bar{a}$, middle: $a_{\rm bg}=0.2\bar{a}$. 
Right: $\Delta B$ for measuring at the CIR (dashed lines) and unit transmission peak (solid lines) for $a_{\rm bg}=9.7\bar{a}$ (red) and $a_{\rm bg}=0.2\bar{a}$ (blue).}
\end{figure*}

{\it Choice of resonance parameters}.
Figure~\ref{fig:prec} displays the precision by means of~Eq.~\eqref{fishers} for the same parameters as in Fig.~\ref{fig:transm}. Quite strikingly, depending on the background scattering length, the precision has a different dependence on the collision energy and it can take the highest values either at the CIR or at the unit transmission peak.
This can be explained by extracting the leading order behavior of Eq.~\eqref{fishers}.
Neglecting finite energy corrections, we obtain that 
\beq
\Delta B = a_{\rm bg}\Delta\left(\frac{1}{p d^2}+\frac{C^2 p}{4}\right)
\eeq
at the CIR and
\beq
\Delta B = \frac{\Delta p d^2}{4a_{\rm bg}}
\eeq
at the unit transmission peak. In both cases $\Delta B$ scales linearly with the resonance width $\Delta$, which gives a natural scale for the detection uncertainty. However, at the CIR a low background scattering length and a certain finite $p$ is preferred, while at the unit transmission peak a high $a_{\rm bg}$ and a very low energy gives better results. These simple bounds are in good agreement with the general formula~\eqref{fishers} and are summarized in Fig.~\ref{fig:precdelta}.

Let us now discuss the realistic conditions for the implementation of the sensor. For measuring at the unit transmission peak, it is preferred to work with high background scattering length which is typically expected e.g. in ion-atom mixtures or between Cs atoms. However, in order to achieve the highest precision for this case, it is required to reduce the collision energy to sub-nK regime. To be able to work at more reasonable temperatures it is better to switch to the CIR and low $a_{\rm bg}$. Here the most promising systems are the ones involving lanthanide atoms such as dysprosium and erbium, which can feature tens of narrow resonances per Gauss along with a few broad resonances that set the local background~\cite{Frish2014,Maier2015,Maykel2015}. This ensures that one can find at least a few resonances with the desired properties. Scattering of lanthanide atoms includes sizeable dipolar contribution~\cite{SupMat} and represents a challenge for a full theoretical description and is the subject of intense investigations~\cite{Petrov2012,Maier2015X}. Finding a resonance with $\Delta\approx 0.01$G in the region where $a_{\rm bg}\approx 0.1\bar{a}$ leads to a precision of the order of $10^{-5}$G (single nanotesla) with a single atom at reasonable energies~$\sim 10$nK.

{\it Discussion}. We proceed with discussing the main potential error sources. The first limiting factor is the finite width of the longitudinal momentum distribution. In the case of measuring at unit transmission, this can impose stringent limits as one has to reduce the energy as much as possible. However, as can be seen from the right panel of Fig.~\ref{fig:precdelta}, for measuring at the CIR $\Delta B$ has a rather broad minimum. In addition, one has to consider fluctuations of the resonance position due to finite energy corrections given by the differential magnetic moment $\delta\mu$, which typically is of the order of several MHz per Gauss~\cite{Chin2010}. This results in uncertainty of the order of $10^{-5}$ G for energy distribution width of 1 nK. This estimation shows that the momentum has to be quite precisely controlled.

Furthermore, the details of the trapping potentials and interatomic interactions can lead to emergence of additional narrow resonances along with a shift of the $s$-wave CIR position. These system-specific effects have to be included, but do not affect the precision bounds~\cite{SupMat}.

In addition, the uncertainty of the estimated magnetic field strength depends on the efficiency of the detector, denoted by $0\leqslant\eta\leqslant 1$.
Let us assume first that we measure whether the atom injected into the tube was transmitted or reflected. Then, the probability of detecting the transmitted (reflected)
atom is given by $\eta P(\pm1|B)$. In such a case, the Fisher information is simply given by $F^{(\mathrm{I})} = \eta F$, where $F$ is the Fisher information for the perfect detectors given by Eq.~\eqref{cfi} and the attainable uncertainty $\Delta B$ is rescaled by a factor $1/\sqrt{\eta}$.

In another scenario one can measure only the transmitted atoms and the reflected atoms are not monitored. In this case, the fact that we do not detect an atom can be due either to reflection or to the detector inefficiency. Therefore, the probability of registering the transmitted atom is $\eta P(+1|B)$, whereas the probability of not detecting this atom is $1-\eta P(+1|B)$. Consequently, the Fisher information is given by
$F^{(\mathrm{II})} = \eta (T'(B))^2/T(1-\eta T(B))$. In the limit $\eta=1$ we recover the result $F^{(\mathrm{II})}=F$ for two perfect detectors.

In the proposed scheme we measure only the number of atoms that were transmitted through
the impurity. It is natural to ask about the maximal attainable precision utilizing a different measurement. To answer this question we refer to the QCRLB,
which provides the lower bound for the precision of any measurement allowed by quantum mechanics.  In~\cite{SupMat}, we show that the measurement we propose
yields a precision of the magnetic field that saturates this bound. As a consequence, a different measurement strategy, preceded optionally by any operation on the state of
the system after the collision, will not improve the precision further. This result can be understood as follows.  After the collision the particle is in a superposition
of being transmitted or reflected with respective probability amplitudes. The modulus and phase of these amplitudes depend on the magnetic field. The measurement we
propose is only capable of determining the moduli of the amplitudes, but the information about the field encoded in the phases is lost.  However, the phases of the
amplitudes are equal and form a common phase factor.  As a consequence, in our situation, the full information about the magnetic field is contained in the moduli of the amplitudes, and, thus, the measurement is optimal~\cite{wasak2016optimal}.


Finally, let us compare the performance of our collisional sensor to other available magnetic field sensing methods. 
At this point it is convenient to take into account that accumulation of the data improves the sensitivity. The scaling with the number of repetitions improves the precision by a factor $1/\sqrt{M}$. Denoting the time for detecting a single collision
event by~$\tau$, during the total time~$t$ of the experiment $M = t/\tau$ repetitions are made.  For reasonable times $\tau$ of the order of a few tens of miliseconds, the
achievable precision scales with $t^{-1/2}$ as $10-100\,\mathrm{pT\,Hz}^{-1/2}$.
The sensor we propose is sensitive to static (dc) magnetic fields, and thus works in a different regime than SQUIDs, trapped ions or NV centers.  Its small radial size of the order of a few tens of nanometers makes it especially useful for probing the local magnetic field directly in the experiments based on cold atoms. Furthermore, with the optical lattice forming the waveguides, the sensor can work as a parallel, multipoint scanning probe capable of measuring local
magnetic fields with a sub-micron resolution limited by the lattice spacing. This configuration can be valuable for measuring the field gradients. The combination of high resolution with nanotesla precision is unique compared to other methods of dc field sensing (see Fig.~4 in \cite{Yang2017} for a detailed comparison).

{\it Conclusions}. We have demonstrated that \mbox{Feshbach} and confinement-induced resonances can make cold collisions useful from a quantum sensing point of view. We identified an optimal measurement scheme in which both reflected and transmitted atoms are monitored after the scattering event. We proved that in this approach the sensitivity of the magnetic
field is maximal and we saturate the Quantum Cram\'er-Rao lower bound. This approach can allow for ultraprecise characterization of Feshbach resonances, overcoming the three-body loss measurements which are sensitive to temperature effects and detailed structure of the three-body bound states. It might find application in precise determination of the residual magnetic fields for improved precision of optical lattice clocks. It would also be interesting to extend the scheme beyond magnetic field
measurements. It is well known that Feshbach resonances can be controlled with external laser and rf fields, making cold collisions a possibly versatile sensor.

This work was supported by the Alexander von Humboldt Foundation, the Polish National Science Center project 2014/14/M/ST2/00015, the cluster of excellence The  Hamburg  Centre  for  Ultrafast  Imaging  of  the  Deutsche  Forschungsgemeinschaft, and the European Union FP7 FET Proactive project DIADEMS (grant N.  611143).

\bibliography{allarticles}
\end{document}